\documentclass{ws-procs10x7}
\usepackage{balance}
\usepackage{graphicx}

\newcolumntype{d}[1]{D{.}{.}{#1}}

\makeindex
\begin{document}

\title{Lattice approach to threshold states}

\author{C.~McNeile}

\address{Department of Physics and Astronomy, The Kelvin Building\\
University of Glasgow, Glasgow G12 8QQ, U.K.
\\
$^*$E-mail: c.mcneile@physics.gla.ac.uk}


\twocolumn[\maketitle\abstract{
I review lattice studies of the hadrons:
$D_s(2317)$, $X(3872)$, and
$Y(4260)$.
}
\keywords{Lattice QCD; Spectroscopy; Heavy quarks.}
]

\section{Introduction}

I review lattice QCD results relevant to the recently discovered
hadrons: $X(3782)$, $Y(4260)$, and $D_s(2317)$, because these
seem, to me at least, to be the most interesting states from the
perspective of solving and understanding
QCD~\cite{Swanson:2006st,Godfrey:2006pd}.

The physical picture behind lattice QCD calculations is that
an interpolating operator creates a hadron in the QCD vacuum.
and
after a specific time interval the hadron is destroyed.
The choice of interpolating operator is particularly
important for hadrons that are thought to be a hybrid
meson or a molecule of hadrons. 

For example, for a $1^{--}$ state in the charmonium system,
possible interpolating operators are
\begin{eqnarray}
O_1 & = &  \overline{c}\gamma_i c  \label{eq:standard} \\
O_2 & = &  \epsilon_{ijk} \overline{c}\gamma_5 F_{j k} c 
\label{eq:hybrid} \\
M_2 & = & (\overline{c} \gamma_i c)  (\overline{q} q) \label{eq:molecule}
\end{eqnarray}
where $F_{j k}$ is the QCD field strength tensor, and $c$ and
$q$ are creation operators for the charm and light quark
respectively.
Operator 
$O_2$ is a hybrid meson operator because it contains excited glue.

 
A critical issue for molecular
interpolating operators, such as $M_2$ in equation~\ref{eq:molecule},
is whether the state is a genuine bound state or two mesons weakly
interacting. One technique that is widely used, was developed by the
Kentucky group~\cite{Mathur:2004jr}, is to study the volume dependence
of certain amplitudes in the calculation.  For non-interacting
scattering states the amplitude is proportional to the inverse of the
volume, but for resonances the amplitude is independent of the volume.

An important issue is how close the parameters of unquenched lattice QCD
calculations are to the QCD in the real world. The MILC
collaboration have pion masses as low as 240 MeV, a dynamical range of
lattice spacings between 0.06 and 0.15 fm, and 2+1 flavours of sea
quarks~\cite{Bernard:2006wx}.  
As reviewed by Schierholz at this conference, due to
algorithm breakthroughs, other collaborations are now doing lattice
calculations with comparable parameters, For example the ETM
collaboration~\cite{Jansen:2006rf} have accurate results at two
lattice spacings with pion masses just under 300 Mev with 2 flavours
of sea quarks.
Unfortunately, the published results on the new heavy hadrons 
use older data sets, that are either
quenched or unquenched with pion masses above 500 MeV.

The effect of the sea quarks could be important for hadrons close to
threshold, such as the $D_s(2317)$ and $X(3872)$.
In an unquenched lattice calculation, a sea quark loop in a
meson has the quark content of $\overline{q_1}\overline{q_s}q_sq_2$. This dynamics is
important for the mixing of tetraquark and quark-antiquark states. Also
this diagram contains the dynamics of two meson decay that are sometimes
included in quark models via coupled channel 
techniques~\cite{Swanson:2006st,Michael:2005kw}

\section{Lattice results for $D_s(2317)$.}


This state was discovered by BaBar and confirmed by CLEO, and
BELLE~\cite{Swanson:2006st,Godfrey:2006pd}.  The quantum numbers are
thought to be $J^{P} = 0^+$. The quark model predictions for the
$J^{P} = 0^+$ strange-charm meson were above the $DK$ threshold. 
The experimental signal for the $D_s(2317)$
was below the $DK$ threshold with a small width.  The closeness of the
mass of the $D_s(2317)$ state to the mass of the $DK$ threshold has
caused some people to speculate that the experimental $D_s(2317)$ is
a hadron molecule.

After the discovery of the $D_s(2317)$,
Bali~\cite{Bali:2003jv} estimated the mass of the lightest
strange-charm $0^{+}$ meson to be 2.57(11) GeV
from lattice QCD and suggested this provided
some evidence for non-$\overline{c}s$ interpretation
of the $D_s(2317)$.
UKQCD, obtained the mass 2404(57) MeV for $0^{+}$
using charm-strange interpolating operators,
and claimed consistency between lattice and 
mass of $D_s(2317)$~\cite{Dougall:2003hv}.
Lin et al.~\cite{Lin:2006vc} recently reported 
the mass of the $D_{s}(0^+)$ to be 2379(40) MeV,
from a quenched lattice QCD calculation at a single 
lattice spacing. 


Another way of determining whether a state is a molecule
or not is to compute the leptonic decay
constant of the state~\cite{McNeile:2006nv,McNeile:2004rf}.
$$
\langle 0 \mid \overline{c} s | D_{0^+} \rangle
= M_{D_{s\,0^+}} f_{D_{s\,0^+}}
\label{eq:decayDEFN}
$$
There are some normalisation issues for leptonic
decay constant that are discussed in~\cite{Herdoiza:2006qv}.
A molecular state would have a small coupling 
to the $\overline{c} s$ operator at the origin.
From partially unquenched QCD 
UKQCD obtained~\cite{Herdoiza:2006qv}
$f_{D_{s\,0^+}}$ = 340(110) MeV.
This is large on the scale relative to the 
pion decay constant and so is inconsistent with the 
$D_{s\,0^+}$ being molecular. The size of the leptonic decay
constant can not discriminate between a localised 4 quark
state and quark anti-quark state~\cite{Herdoiza:2006qv}. 
Other uses of $f_{D_{s\,0^+}}$ in phenomenology are
discussed in~\cite{Herdoiza:2006qv}. 


After the discovery of the $D_s(2317)$, UKQCD studied
the spectrum of the $B_s$ states 
using unquenched lattice QCD~\cite{Green:2003zz}. 
Results for the masses of four L=1 strange-bottom mesons were
presented, as well as arguments for all four states to have narrow widths.
For example, UKQCD~\cite{Green:2003zz} obtained
$M(B_{s0}) - M(B_s)$ = $386 \pm 31$ MeV that is under the 
$B_s \;K$ threshold.
One of the results from ~\cite{Green:2003zz}:
$M(B_{s2}) - M(B_s)$ = $534 \pm 52$ MeV,
can be compared against
the preliminary result from D0~\cite{Catastini:2006sq}
of $469 \pm 1.4 \pm 1.5$ MeV.

UKQCD used unquenched lattice QCD to
compute the decay width of 160 MeV for the lightest P-wave
B meson to decay to the S-wave B meson and a 
pion~\cite{McNeile:2004rf}.
Also
an effective hadronic coupling for the decay of the lightest 
P-wave $B_s \rightarrow BK$
was found to be of similar size to the coupling for the
decay $K(1412)$ to $K \pi$. Since, the $K(1412)$ is not thought to
be molecular, this is additional
evidence that the lightest P-wave $B_s$ meson is not
molecular.

\section{Lattice results for $X(3872)$.}

The $X(3872)$ was first discovered by 
Belle~\cite{Swanson:2006st,Godfrey:2006pd}. The
mass is $3872.0 \pm 0.6 \pm 0.5$ MeV and the
width is less than 2.3 MeV~\cite{Swanson:2006st,Godfrey:2006pd}.
The $X(3872)$ is thought to have $J^{PC}$ = $1^{++}$ quantum
numbers.  The assignment $J^{PC}$ = $2^{-+}$ 
for the  $X(3872)$ 
has not been ruled out~\cite{Swanson:2006st,Godfrey:2006pd}.
The mass of $X(3872)$ is very close to 
the $D^0 \overline{D^\star}^0$ threshold and this 
motivated the suggestion that the  $X(3872)$ is a molecule
(other possibilities are reviewed in~\cite{Swanson:2006st}).


Quenched lattice calculations~\cite{Chen:2000ej} 
of the charmonium spectrum
find the first excited state with $J^{PC}$ = $1^{++}$ above
4.00(8) GeV (statistical errors only) using quark and anti-quark
interpolating operators.

Chiu and Hsieh~\cite{Chiu:2006hd} have used quenched lattice QCD
to study the $1^{++}$
state in charmonium using molecular and diquark-antidiquark operators.
They see a state at $3890 \pm 30$ MeV~\cite{Chiu:2006hd} that has the 
expected volume dependence for a resonance. One concern about the 
results of Chiu and Hsieh~\cite{Chiu:2006hd} is that their quark masses
are large ($m_c a $ = 0.8) in lattice units. This could mean that
the systematic errors due to the non-zero lattice spacing  are potentially large.
Using the same lattice setup Chiu et al.~\cite{Chiu:2005ue}
computed the $f_D$ and $f_{D_s}$ decay constants and obtained
good agreement with the recent experimental measurements 
by CLEO-c, so this is a crosscheck on their systematic errors.

\section{Lattice results on $Y(4260)$.}

The $Y(4260)$ (with $J^{PC}$ = $1^{--}$) was first seen by BaBar and
has been confirmed by
CLEO~\cite{Swanson:2006st,Godfrey:2006pd}. Although there are many
suggestions for the quark and glue content of the $Y(4260)$, perhaps
the most popular one is that the state is a non-exotic hybrid meson (this
still needs confirming of
course)~\cite{Swanson:2006st,Godfrey:2006pd}.

\begin{figure}
\includegraphics[%
  scale=0.3,
  angle=270,
  origin=c,clip]{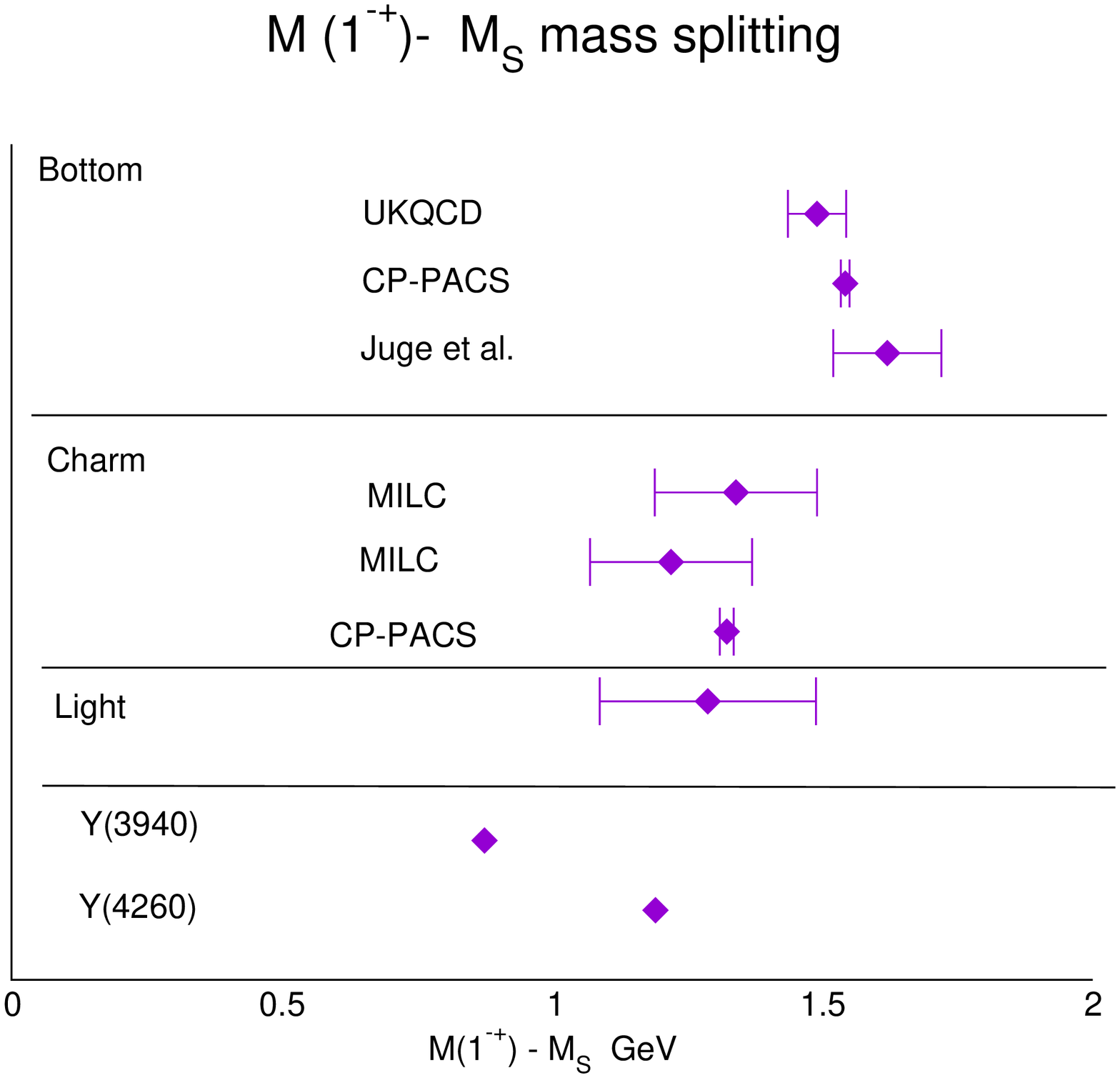}
\label{fig:hybridMASS}
\end{figure}

There have been a lot of lattice calculations that studied 
the exotic charmonium 
meson with
$J^{PC}$ = $1^{-+}$.
Although $J^{PC}$ = $1^{--}$ for the $Y(4260)$ hadron, the 
mass of the $1^{-+}$ state gives some indication of the hybrid 
excitation energy.
In figure~\ref{fig:hybridMASS}, I
plot the mass difference between the $1^{-+}$ states and the S-wave
states for both heavy and light quarks using data
from~\cite{Juge:1999ie,Michael:2003ai}, as well as the experimentally
determined masses of the $Y(3940)$ and $Y(4260)$ states.
This shows that the 
mass of the $Y(4260)$ state is close to the mass
of the $1^{-+}$ masses, but the mass of the $Y(3940)$ is
too low~\cite{Swanson:2006st}.

In the heavy quark limit the decay width of  the 
exotic $1^{-+}$ meson has been computed via lattice
QCD~\cite{McNeile:2002az}. 
The excited potential de-excites to the standard 
heavy quark potential (Coulomb + linear) via the 
emission of light quark-antiquark pair. This transition was
used to estimate the widths for 
$1^{-+}  \rightarrow \chi_b S $  as $\sim$ 80 MeV
and the width for $1^{-+}  \rightarrow \chi_b \eta $ 
to be less than 1 MeV,
where $S$ is a scalar $0^{++}$ meson.
The message from
the lattice gauge theory 
calculation in~\cite{McNeile:2002az} is that 
the decay width of hybrid mesons to states that include light flavour
singlet mesons could be sizeable.

Luo and Liu~\cite{Luo:2005zg} studied
the non-exotic hybrid mesons in charmonium using
a quenched lattice QCD calculation. The masses of the ground and
excited states that coupled to the $1^{--}$ operator in 
equation~\ref{eq:standard} were 3.094(18) GeV (close to $J/\psi$)
and 3.682(81) GeV (close  to $\psi(2S)$). 
The masses of the ground and
excited states that coupled to the $1^{--}$ hybrid meson operator in 
equation~\ref{eq:hybrid} were 3.099(62) GeV (close to $J/\psi$)
and 4.379(149) GeV (close  to $Y(4260)$). 
My main
criticism of the work is that they use multi-exponentials fit models
to fit single channel correlators. They used reasonable techniques, 
but that this type of fitting is still hard to do.

Chiu and Hsieh~\cite{Chiu:2005ey} used
hybrid and molecular operators 
(equations~\ref{eq:standard}, ~\ref{eq:hybrid},and ~\ref{eq:molecule} )
and additional operators to study the $1^{--}$ hadron
in charmonium using a quenched lattice QCD calculation at
a single lattice spacing. For the first excited state
of the hybrid $1^{--}$ operator they obtain 
the mass $4501(178)(215)$ MeV in reasonable agreement
with the calculation of Luo and Liu~\cite{Luo:2005zg}
(whose result is reduced by taking the continuum limit).
Using a molecular operator of the form 
$(\overline{q}c \overline{c}q)$,
Chiu and Hsieh~\cite{Chiu:2005ey} obtained the
mass $4238 \pm 31 $ MeV and they used the Kentucky 
volume
method~\cite{Mathur:2004jr} to show that the state
was a resonance. Hence Chiu and Hsieh~\cite{Chiu:2005ey}
favour a molecular interpretation of the $Y(4260)$.

Burch at al.~\cite{Burch:2001tr,Burch:2003zf} 
have studied the explicitly 
mixing between the hybrid operator
and $\overline{q}q$ operator for the $1^{--}$ state
using NRQCD in a quenched lattice QCD
calculation. The $\frac{\sigma B}{2M_{Q}a}$ term
in the NRQCD Lagrangian is the one that mixes the hybrid and 
$\overline{q}q$ operators to the order that they work.
For the ground state $Y$ they obtain the mixture
$$
\mid Y \rangle =  
0.99826(6) \mid \overline{Q} Q \rangle 
-0.059(1) \mid \overline{Q} Q g \rangle 
$$
so the hybrid $\mid \overline{Q} Q g \rangle$ contribution to
the ground heavy-heavy $1^{--}$ state is small.
A similar calculation for the first excited $1^{--}$
hadron would be useful to help understand the $Y(4260)$,
unfortunately the NRQCD expansion is not very reliable for charmonium.

\section{Conclusions}

My first "no-brainer" conclusion is that the above lattice
calculations need to be repeated with modern unquenched lattice QCD
data sets.  It is particularly important to study the effect of sea
quarks on the tetraquark/molecular 
interpolating operators. In quenched QCD, it
seems that tetraquark/molecular 
and $\overline{q}q$ operators couple to distinct
states, however sea quarks will in principle cause these two type of
operators to mix.  The molecular versus quark-antiquark picture is
also an issue for light hadrons such as the $a_0(980)$.  Two recent
lattice calculations~\cite{McNeile:2006nv,Mathur:2006bs} disagree on
the quark content of the $a_0(980)$.

In QCD there are also glueball interpolating
operators. The potential effect of glueball dynamics on
vector and pseudoscalar states in charmonium is
discussed in~\cite{McNeile:2004wu}.


\end{document}